\documentclass[aps,prd,preprint,superscriptaddress,tightenlines,nofootinbib]{revtex4}

\usepackage{graphicx}
\usepackage{dcolumn}
\usepackage{bm}

\begin{document}

\preprint{\tighten\vbox{\hbox{\hfil CLNS 04/1882}
                        \hbox{\hfil CLEO 04-07}
}}

\title{Search for the Lepton-Flavor-Violating Leptonic $B$
Decays
$B^0 \rightarrow \mu^\pm \tau^\mp$ 
and 
$B^0 \rightarrow e^\pm \tau^\mp$}

\author{A.~Bornheim}
\author{E.~Lipeles}
\author{S.~P.~Pappas}
\author{A.~J.~Weinstein}
\affiliation{California Institute of Technology, Pasadena, California 91125}
\author{R.~A.~Briere}
\author{G.~P.~Chen}
\author{T.~Ferguson}
\author{G.~Tatishvili}
\author{H.~Vogel}
\author{M.~E.~Watkins}
\affiliation{Carnegie Mellon University, Pittsburgh, Pennsylvania 15213}
% ...
\collaboration{CLEO Collaboration} 
\noaffiliation

\date{5 August 2004}

\begin{abstract}
We have searched a sample of 9.6 million $B \bar B$ events for 
the lepton-flavor-violating leptonic $B$ decays, 
$B^0 \rightarrow \mu^\pm \tau^\mp$ and $B^0 \rightarrow e^\pm \tau^\mp$.
The $\tau$-lepton was detected through the decay modes 
$\tau\rightarrow\ell\nu\bar\nu$, where $\ell = e, \mu$.
There is no indication of a signal, and we obtain the 90\% confidence level 
upper limits
${\cal B}(B^0 \rightarrow \mu^\pm \tau^\mp) < 3.8 \times 10^{-5}$ and
${\cal B}(B^0 \rightarrow e^\pm \tau^\mp) < 1.1 \times 10^{-4}$.

\end{abstract}

\maketitle


We report results of a search for two lepton-flavor-violating leptonic
decays of $B$ mesons: $B^0 \rightarrow \mu^\pm \tau^\mp$ 
and $B^0 \rightarrow e^\pm \tau^\mp$.
These modes are forbidden in the conventional 
Standard Model by the lepton-flavor 
conservation law. However, they are predicted to occur in many theories
``beyond the standard model,'' for example multi-Higgs-boson extensions, 
theories with leptoquarks, supersymmetric models without R-parity, and
Higgs-mediated decay in supersymmetric seesaw models \cite{theory}. 
The recent discovery of neutrino oscillation, while not leading to 
predictions of observable rates for lepton-flavor-violating decays, 
nonetheless heightens interest in them \cite{neutrino}.
The decays we searched for
involve both third generation quarks and third generation leptons.
Decays of this variety have been less extensively searched for than those 
involving only first or second generation quarks or leptons. 
Discovery of such 
decays at levels of our sensitivity 
would be clear evidence of physics beyond the Standard Model. 
Currently the best limits on the branching
fractions are ${\cal B}(B^0 \rightarrow \mu^\pm \tau^\mp) < 8.3\times 10^{-4}$,
and ${\cal B}(B^0 \rightarrow e^\pm \tau^\mp) < 5.3\times 10^{-4}$, at
90\% confidence level \cite{CLEO-old}.

    The data used in this analysis were taken with the CLEO
detector \cite{detector} at the Cornell Electron Storage Ring (CESR), a 
symmetric
$e^+ e^-$ collider operating in the $\Upsilon({\rm 4S})$ resonance region. The
data sample consists of 9.2 ${\rm fb}^{-1}$ at the resonance, corresponding to
9.6 million $B \bar B$ events,
and 4.5 ${\rm fb}^{-1}$ at a center-of-mass energy 60 MeV below the resonance.
The sample below the resonance provides information on the background from
continuum processes $e^+ e^- \rightarrow q \bar q,\ q = u,d,s,c$, 
and from two-photon fusion processes 
$e^+ e^- \rightarrow e^+ e^-\gamma^*\gamma^*, 
\gamma^*\gamma^*\rightarrow X$ ($\gamma^*$ a virtual photon). 
We scale the off-resonance yields
by 1.99, the luminosity ratio divided by the c.m. energy-squared ratio,
and subtract them from on-resonance yields to obtain $B\bar B$ yields.

Summing over $\mu(e)^+\tau^-$ and $\mu(e)^-\tau^+$,
we search for 
$B^0\rightarrow \mu(e)^\pm \tau^\mp$ with the $\tau$ lepton detected via the
$\tau \rightarrow e \nu \bar\nu$ and $\tau \rightarrow \mu \nu \bar\nu$
decay modes. In this article,
$\ell$ denotes the primary lepton from the signal $B$ and $\ell^\prime$
denotes the secondary lepton from $\tau$. ($\ell$, $\ell^\prime$) denotes
$B^0\rightarrow \ell^\pm \tau^\mp$, $\tau \rightarrow \ell^\prime \nu \bar\nu$.
We have four modes to analyze; ($\mu$, $e$), ($\mu$, $\mu$), ($e$, $e$),
and ($e$, $\mu$).


    Muons are identified by their ability to penetrate the iron return
yoke of the magnet: at least five (three) interaction lengths of material for 
the primary (secondary) muon. Electrons are identified by shower energy to 
momentum ratio ($E/P$), track-cluster matching, $dE/dx$, and shower shape.

    In the rest frame of the signal $B$, the primary lepton is monoenergetic, 
with momentum 2.34 GeV/c. In the lab frame (the $\Upsilon$(4S) rest frame),
this is smeared, and ranges from 2.2 to 2.5 GeV/c. 
We require that the primary lepton candidate have momentum in that range.
We require that the
secondary lepton, from $\tau$, be greater than 0.6 (1.0) GeV/c for $e$ ($\mu$).
We ``measure'' the 4-momentum of the neutrino
pair as the missing visible 4-momentum in the event :
$E_{\nu\bar\nu} = 2E_{beam} - \Sigma E_i$, 
$\vec P_{\nu\bar\nu} = - \Sigma \vec P_i$, 
where sums are over all observed (charged and neutral) particles.

    We define two neural net variables.
$NN_{B\bar B}$ is a neural net variable used to suppress backgrounds from
$B\bar B$ decays. We calculate three inputs: beam-constrained mass
$\sqrt{E^2_{beam}-P^2_{cand}}$, 
$\Delta E \equiv E_{cand} - E_{beam}$, where $P_{cand}$ ($E_{cand}$) 
is the momentum (energy) of the $B$ candidate, 
and $\cos \theta_{\ell B}$
(the cosine of the angle between the momenta of primary lepton and
$B$ candidate). We feed them into a neural net and train it with signal and
$B\bar B$ Monte Carlo simulations for each mode. 
$NN_{cont}$ is a neural net variable to suppress 
backgrounds from continuum. We calculate five inputs: $R_2$ 
(the ratio of second and zeroth Fox-Wolfram moments \cite{Fox-Wolfram} of the
event), $S$ (the sphericity), thrust of the event, $\cos \theta_{tt}$ 
(the cosine of the angle between the $\vec p_\ell - \vec p_{\ell^\prime}$ and 
the thrust axis of the rest of the event), and 
$\cos\theta_{\vec p_{\nu\bar\nu}, \vec p_\ell+ \vec p_{\ell^\prime}}$ 
(the cosine of the angle
between the momenta of neutrino pair and lepton pair), 
then feed them into a neural net and train it with
signal and continuum Monte Carlo for each mode. 
The nominal neural net range is from 0.0 to 1.0.
We cut in the 2D space defined by $NN_{B\bar B}$ and $NN_{cont}$, 
requiring $NN_{B\bar B} > NN^{cut}_{B\bar B}$, $NN_{cont} > NN^{cut}_{cont}$
and also
$$ \frac{(NN_{B\bar B} - NN^{cut}_{B\bar B})}{(1-NN^{cut}_{B\bar B})}
+ \frac{(NN_{cont}-1)}{(1-NN^{cut}_{cont})} > 0.$$

    We define two $\tau$-mass variables. The first is the 
conventionally defined invariant 
mass of the reconstructed $\tau$, $M_{\ell^\prime\nu\bar\nu} \equiv 
\sqrt{(E_{\ell^\prime} + E_{\nu\bar\nu})^2 
- (\vec p_{\ell^\prime} + \vec p_{\nu\bar\nu})^2}$. 
The second $\tau$-mass variable makes use of the fact that, 
with perfect measurements of all quantities, $\Delta E = 0$, 
and hence we can use $E_{beam} - E_\ell - E_{\ell^\prime}$ for
$E_{\nu\bar\nu}$, 
yielding $M_{\ell^\prime \nu\bar\nu, \Delta E = 0} \equiv 
\sqrt{(E_{beam} - E_\ell)^2 
- (\vec p_{\ell^\prime} + \vec p_{\nu\bar\nu})^2}$.
We further define $\Delta M_\tau \equiv M_{\ell^\prime\nu\bar\nu} -M_\tau$
and $\Delta M_{\tau, \Delta E =0} \equiv 
M_{\ell^\prime\nu\bar\nu, \Delta E =0} -M_\tau$, where $M_\tau$ is the
nominal $\tau$ mass, 1777 MeV.

    By examining the angular distribution of electrons, positrons, and missing 
momentum, in off-resonance data in a $|\vec p_\ell|$ sideband region
($2.0 < |\vec p_\ell| < 2.2$ GeV/c and $2.5 < |\vec p_\ell| < 2.7$ GeV/c),
we see clear evidence of the two-photon-fusion process. There are sharp
peaks in the forward directions for electrons and positrons
(``forward'' being the direction of the beam particle of the same charge).
Also, the missing momentum peaks sharply in the opposite direction from 
a detected $e^+$ or $e^-$, indicating an $e^-$ or $e^+$ lost down the beam 
pipe. By eliminating events with 
$|\cos\theta_{miss}| > 0.90$ (0.95 for ($\mu$, $\mu$)),
we considerably reduce this background.

\begin{figure}[h]
\begin{center}
\includegraphics*[width=0.4\textwidth]{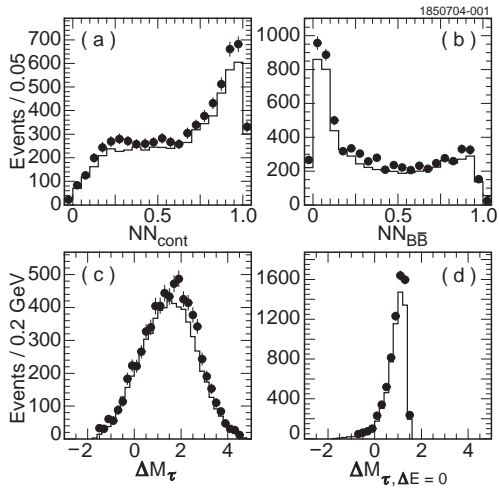}
\caption{ (a) $NN_{cont}$, (b) $NN_{B\bar B}$, 
(c) $\Delta M_\tau$, and (d) $\Delta M_{\tau, \Delta E = 0}$ 
distributions, for the comparison of
$B\bar B$ Monte Carlo (histogram) vs 
off-resonance-subtracted on-resonance data (points)
in the $\vec p_\ell$ sideband region, for the ($\mu$, $e$) mode.}
\label{figureI}
\end{center}
\end{figure}

\begin{figure}[h]
\begin{center}
\includegraphics*[width=0.4\textwidth]{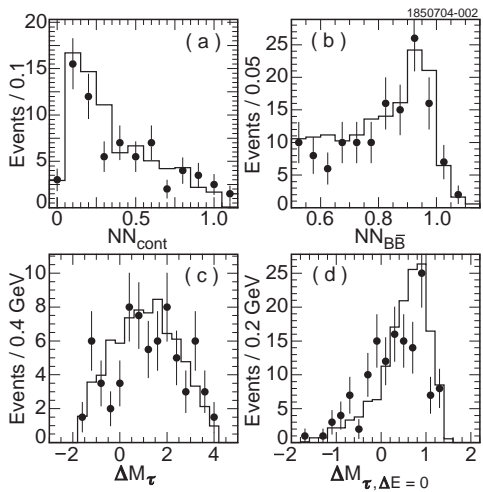}
\caption{ (a) $NN_{cont}$, (b) $NN_{B\bar B}$, 
(c) $\Delta M_\tau$, and (d) $\Delta M_{\tau, \Delta E = 0}$ 
distributions, for the comparison of
continuum Monte Carlo ($e^+ e^- \rightarrow q \bar q,\ q = u,d,s,c$: histogram)
vs off-resonance data (points)
in the $\vec p_\ell$ sideband region, for the ($\mu$, $e$) mode.}
\label{figureII}
\end{center}
\end{figure}

	We compare Monte Carlo samples with data using the 
$|\vec p_\ell|$ sideband region defined above. In Fig.~\ref{figureI},
we show distributions in $NN_{cont}$, $NN_{B\bar B}$, 
$\Delta M_\tau$, and $\Delta M_{\tau, \Delta E =0}$, for 
off-resonance-subtracted on-resonance data and absolutely normalized $B\bar B$
Monte Carlo. Agreement is good. In Fig.~\ref{figureII}, we show 
distributions for the same variables for off-resonance data and
absolutely normalized continuum ($e^+ e^- \rightarrow q \bar q,\ q = u,d,s,c$)
Monte Carlo. We have {\it not} included a Monte Carlo for the inclusive 
multi-hadronic two-photon-fusion process, 
lacking a trustworthy simulation of this process.
For ($\mu$, e), shown in Fig.~\ref{figureII}, agreement is good,
indicating that the remaining contribution from two-photon fusion is
small. For ($e$, $e$), not shown, data exceeds continuum Monte
Carlo, indicating a sizeable remaining contribution from two-photon fusion.

We measure the ratio of data to  
Monte Carlo yields in the $|\vec p_\ell|$ sideband region,
denoting by ${\cal R}_{B\bar B}$ the ratio of off-resonance-subtracted
on-resonance data to $B\bar B$ Monte Carlo, and by 
${\cal R}_{cont}$ the ratio of off-resonance data to
continuum Monte Carlo.
${\cal R}$'s are measured with loose selection criteria applied:
$NN_{cont} > 0.5$, $NN_{B\bar B} > 0.5$, 
$|\cos\theta_{miss}|<$ 0.9 (0.95 for ($\mu$, $\mu$)),
and $|\Delta M_\tau| < $ 2.0 GeV.
Values so obtained are given in 
Table~\ref{table}. 
One sees that ${\cal R}_{B\bar B}$ differs little from 1.0, 
while ${\cal R}_{cont}$ is less well behaved, particularly
for the ($e$, $e$) mode, indicating that the two-photon fusion
background is present. To estimate backgrounds in the 
signal $|\vec p_\ell|$ region correctly, we scale the 
signal-region yields from Monte Carlo by ${\cal R}$.
Because we perform a direct subtraction of off-resonance
data, the accuracy of the continuum background prediction is not critical
for our results.

\begin{table}[h]
\begin{center}
\caption{The rows of $NN^{cut}_{B\bar B}$, $NN^{cut}_{cont}$, 
$\Delta M_\tau$ (GeV), and
$\Delta M_{\tau,\Delta E =0}$ (GeV) indicate the optimized selection criteria
 for each mode.
${\cal R}_{B\bar B}$ and ${\cal R}_{cont}$ are measured ratios between
data yields and Monte Carlo with loose selection criteria
in the $|\vec p_\ell|$ sideband
region. $N_{ON}$($N_{OFF}$) is the number of observed events 
satisfying the optimized selection criteria in the 
signal region of $|\vec p_\ell|$ from on(off)-resonance data samples.
$N_{obs}$ is the number of observed events from
off-resonance-subtracted on-resonance data, $N_{ON}-$1.99$N_{OFF}$. 
$N_{B\bar B}$ is the $B\bar B$ background estimate from Monte Carlo,
scaled by ${\cal R}_{B\bar B}$; in the absence of signal it should be
comparable with $N_{obs}$.
0.5$N_{cont}$ is (1/1.99) times the 
continuum background estimate from Monte Carlo, scaled by ${\cal R}_{cont}$,
which should be comparable with $N_{OFF}$.
$\epsilon$ is the signal detection efficiency including $\tau$ decay branching
fraction. $BR$ $UL$ is the branching ratio upper limit at 90\% confidence level
with systematic error considered.
\label{table}}

\begin{tabular}{c|cccc}
($\ell$, $\ell^\prime$) & ($\mu$, $e$)& ($\mu$, $\mu$) 
& ($e$, $e$) & ($e, \mu$)  \\\hline
&\multicolumn{4}{c}{} \\
$NN^{cut}_{B\bar B}$ &0.725&0.875&0.675&0.825\\
$NN^{cut}_{cont}$    &0.700&0.775&0.700&0.475\\
--2.0$<$$\Delta M_\tau $  &$<$2.00&$<$1.40&$<$1.50&$<$1.40\\
--2.0$<$$\Delta M_{\tau,\Delta E =0}$&$<$0.25&$<$0.25&$<$0.30&$<$0.25\\
&\multicolumn{4}{c}{} \\
${\cal R}_{B\bar B}$&1.21$\pm$0.06&1.06$\pm$0.07
&1.04$\pm$0.07&0.94$\pm$0.07\\
${\cal R}_{cont}$   &1.03$\pm$0.27&1.52$\pm$0.32
&4.30$\pm$0.95&0.64$\pm$0.38\\
&\multicolumn{4}{c}{} \\
$N_{ON}$   & 19   & 10   & 28   & 6  \\ 
 $N_{OFF}$  &   2   &   3   &   7  	&  0  \\
 $N_{obs}$  &  15.0 $\pm$ 5.2 &  4.0$\pm$4.7 
&  14.0 $\pm$7.5 &  6.0 $\pm$ 2.4\\
&\multicolumn{4}{c}{} \\
 $N_{B\bar B}$  & 23.7$\pm$2.7    & 9.0$\pm$1.4    
& 11.6 $\pm$1.4    & 5.1 $\pm$0.8 \\
 0.5$N_{cont}$ & 1.8$\pm$0.6 & 0.4$\pm$0.2 
& 3.1$\pm$1.0 & 0.5$\pm$0.3\\
&\multicolumn{4}{c}{} \\
$\epsilon$ (\%)   & 1.57 & 0.63 & 0.96 & 0.58 \\ 
&\multicolumn{4}{c}{}\\
$BR$ $UL$($10^{-4}$) & 0.55& 0.87 &1.64 & 1.46 \\
\end{tabular}
\end{center}
\end{table}

We optimized our selection criteria on 
$NN_{B\bar B}$, $NN_{cont}$, $\Delta M_\tau$,
and $\Delta M_{\tau, \Delta E =0}$
to obtain the best upper limit
when the true branching fraction is zero.
This optimization procedure made use of signal and background Monte Carlo
samples, and scaled the background samples by ${\cal R}_{B\bar B}$
or ${\cal R}_{cont}$ as described above. 
The optimized selection criteria, found separately for each mode, are  
shown in Table~\ref{table}.

    The number of events satisfying all selection criteria
is shown, for each mode, in Table~\ref{table},  
along with the background estimate.  We 
find 15 ($\mu$, $e$) candidates, with 23.7 expected from background; 
we find 4 ($\mu$, $\mu$) candidates, with 9.0 expected from background; 
we find 14 ($e$, $e$) candidates, with 11.6 expected from background; 
we find 6 ($e$, $\mu$) candidates, with 5.1 expected from background. 
Thus there are a total of 39
events with 49.4 expected from background. The probability that a true mean of
49.4 will give rise to a yield of 39 or more events is 93\%.  
With no indication of signal, we obtain the branching fraction upper limits.

    We calculate upper limits at 90\% confidence level. There is some 
probability of observing the off-resonance-subtracted on-resonance yield
that we do observe, or less, if the branching fractions for
$B^0\rightarrow\mu^\pm\tau^\mp$ and $B^0\rightarrow e^\pm\tau^\mp$
are zero. We take the 90\% confidence level
upper limit to be that value of the branching fraction which reduces the
above-mentioned probability by a factor of 10. The ingredients needed for the 
calculation are: (1) the observed off-resonance-subtracted on-resonance
yield; (2) the true mean for the background contribution from 
$B\bar B$ processes, and (3) less critically, the true mean of the 
background contribution from non-resonance processes. 
To allow for the uncertainty 
in the background estimates, 
we changed ${\cal R}_{B\bar B}$ and ${\cal R}_{cont}$ in the
unfavorable directions by 1$\sigma$, {\it i.e.} $-1 \sigma$ for
${\cal R}_{B\bar B}$ and $+1 \sigma$ for ${\cal R}_{cont}$.

    We use Monte Carlo simulation to determine the efficiency for detecting 
the signal modes. The decays $B \rightarrow \mu(e)^\pm \tau^\mp$ 
are generated with the $\tau$ lepton unpolarized. 
For a $\tau$ lepton polarization as 
given by $V-A$, the secondary lepton is boosted (has its
average lab energy increased), which in turn increases the efficiency.
For the opposite polarization, as given by $V+A$, the secondary lepton
is deboosted, and the efficiency is lowered. The fractional changes in
efficiency, averaged over the four modes, are $+$11\% for 
$V-A$, $-$8\% for $V+A$. Our upper limits are quoted for unpolarized $\tau$'s.


    Systematic errors are of two varieties -- those on the estimate of
signal detection efficiencies, and those on the estimate of backgrounds.  The
dominant 
contributors to the former are lepton identification efficiency uncertainties 
(contributing $\pm$3.5\% per lepton, relative, in the efficiency), and
missing-four-vector-simulation uncertainties ($\pm$5.4\%), 
giving a relative uncertainty in the overall efficiency of
$\pm$7.4\% for ($e$, $\mu$) and ($\mu$, $e$), and
$\pm$8.9\% for ($e$, $e$) and ($\mu$, $\mu$).
The background uncertainties are handled by varying the 
${\cal R}_{B\bar B}$ and ${\cal R}_{cont}$ as mentioned above.
The errors shown on the backgrounds in Table~\ref{table}
include statistical and systematic errors.

    There is no universally agreed-upon procedure for including systematic
errors in upper-limit estimates.  We conservatively vary the background by
1.0 standard deviations, and decrease the efficiency by 1.0 standard
deviations for each mode and the results are as shown in Table~\ref{table}.   

To combine the results from two leptonic modes, 
$\tau\rightarrow e\nu\bar\nu$ and $\tau\rightarrow \mu\nu\bar\nu$,
we simply add the yields, add the backgrounds, and add the efficiencies.

In this way we obtain our final results:

$${\cal B}(B \rightarrow \mu^\pm \tau^\mp) < 3.8 \times 10^{-5}\ ,$$
$${\cal B}(B \rightarrow e^\pm \tau^\mp) < 1.1 \times 10^{-4}\ ,$$
\noindent both at 90\% confidence level.  These results are significant
improvements over previously published limits \cite{CLEO-old}.

    In summary, we have searched for the decays 
$B \rightarrow \mu^\pm \tau^\mp$ and $B \rightarrow e^\pm \tau^\mp$.  
We find no indication of a signal,
and obtain upper limits on the branching fractions.\\
\indent We gratefully acknowledge the effort of the CESR staff 
in providing us with
excellent luminosity and running conditions.
This work was supported by 
the National Science Foundation,
the U.S. Department of Energy,
the Research Corporation,
and the 
Texas Advanced Research Program.

\end{document}